\documentclass[aps,prl,a4paper,preprintnumbers,twocolumn,floatfix,showpacs,nofootinbib]{revtex4}
\usepackage{amsmath,amsfonts,amssymb,mathrsfs}
\usepackage{graphicx}
\usepackage{subfigure}
\usepackage{enumerate}
\usepackage{color}
\def\be{\begin{equation}}
\def\ee{\end{equation}}
\def\bea{\begin{eqnarray}}
\def\eea{\end{eqnarray}}
\def\nn{\nonumber}

\begin{document}
\title{Black holes in scalar-tensor gravity}

\author{Thomas P.\ Sotiriou$^{1}$ and Valerio Faraoni$^{2}$   
}
\affiliation{$^1$SISSA--ISAS, Via Bonomea 265, 34136, Trieste, 
Italy {\rm and} INFN, Sezione di Trieste, Italy\\
$^{2}$Physics Department and STAR Research Cluster, 
Bishop's University, 2600 College Street, Sherbrooke, Qu\'ebec, 
Canada J1M 1Z7 
}
\begin{abstract}
Hawking has proven that black holes which are stationary as the
endpoint 
of  gravitational collapse in Brans--Dicke theory (without a 
potential) are no different than in general relativity.  We 
extend this proof to the much more general class of  
scalar-tensor and $f(R)$  gravity theories, without assuming any symmetries 
apart from stationarity.
\end{abstract}
\maketitle

In general relativity, spacetime singularities are inevitable, as 
established long ago by the celebrated singularity theorems of 
Hawking and Penrose \cite{Penrose:1964wq,Hawking:1969sw}. It 
seems that, generically, spacetime singularities resulting from 
the collapse of concentrated distributions of mass-energy are 
cloaked by horizons, resulting in black holes 
\cite{Penrose:1969pc}. Additionally, these black holes, provided 
that they are stationary as end-states of collapse, 
also 
have to 
be axisymmetric \cite{hawking1} and are, therefore, rather 
simple objects described by the Kerr--Newman line element.

One is tempted to ask whether black holes in gravity theories 
other than general relativity will also share this last property 
or whether they will be significantly different. This question 
becomes more pertinent if one considers the rather compelling 
indications  that, when one attempts to formulate a quantum 
version of relativistic gravity, one is forced to introduce 
elements foreign to general relativity, such as extra fields  
coupling explicitly to the curvature or higher order 
curvature corrections, which amounts to introducing extra 
degrees of freedom as well. Theories 
with non-minimally coupled scalar fields are typical examples of 
low-energy effective actions for quantum gravity models. 
Additionally,  scalar fields coupled minimally or 
non-minimally to gravity have been extensively studied in 
recent years as potential models of dark energy  
\cite{AmendolaTsujikawabook} ($f(R)$ gravity theories, see 
Refs.~\cite{Sotiriou:2008rp,review2} for reviews,  can ultimately  
be reduced to particular scalar-tensor theories  
\cite{Teyssandier:1983zz,Barrow:1988xi,Wands:1993uu}).

The prototypical scalar-tensor alternative to general relativity 
is Brans-Dicke theory \cite{BransDicke}, which contains a scalar 
field mimicking the dilaton of string theories, and is described 
by the action

\be
\label{bdaction}
S_{\rm BD}=\int d^4x \sqrt{-\hat{g}} \left(\varphi \hat{R}-\frac{\omega_0}{\varphi} \hat{\nabla}^\mu \varphi \hat{\nabla}_\mu \varphi+L_m(\hat{g}_{\mu\nu},\psi)\right)\,,
\ee
where $\hat{g}$ is the determinant and $\hat{R}$ is the 
Ricci scalar of the metric $\hat{g}_{\mu\nu}$,  
$\hat{\nabla}_\mu$ denotes the corresponding covariant 
derivative, $L_m$ is the matter Lagrangian, and $\psi$ 
collectively denotes the matter fields.  It is implicitly assumed 
that the matter fields couple minimally to
$\hat{g}_{\mu\nu}$. This  makes $\hat{g}_{\mu\nu}$ the {\em 
Jordan frame metric} (per definition). It has been shown by 
Hawking in 1972 that black holes which are the 
endpoint of  collapse can be solutions 
of Brans--Dicke  theory if and only if they are also 
solutions of general relativity \cite{hawking2}.

Brans--Dicke theory belongs to a more general class of theories, 
dubbed scalar-tensor theories of gravity, whose action has the 
form
\bea
\label{staction}
S_{\rm st}&=&\int d^4x \sqrt{-\hat{g}} \Big(\varphi 
\hat{R}-\frac{\omega(\varphi)}{\varphi} \hat{\nabla}^\mu \varphi 
\hat{\nabla}_\mu \varphi\nn\\&&\qquad\qquad\qquad-V(\varphi) 
+L_m(\hat{g}_{\mu\nu},\psi)\Big)\,.
\eea
That is, with respect to Brans--Dicke theory, $\omega_0$ has 
been turned into an arbitrary function of $\varphi$ and a 
potential for $\varphi$ has been added. 
One could think of generalizing the  action further by turning 
the coupling to $\hat{R}$ into a general  function of 
$\varphi$ as well, but this essentially amounts to a 
redefinition of the scalar and does not really lead to a new 
theory (provided that the redefinition is regular with 
regular inverse). One could also think of coupling the matter 
minimally to some conformal metric $\Omega^2(\varphi) 
\, \hat{g}_{\mu\nu}$. But, again, this would just mean 
expressing  the same theory in a different conformal frame, not 
actually considering another theory (provided that the conformal 
factor remains regular). In conclusion,  {\em in the Jordan 
frame} the most general scalar-tensor theory  is characterized by 
two arbitrary functions of $\varphi$, $\omega(\varphi)$  and 
$V(\varphi)$.\footnote{See Refs.~\cite{Flanagan:2003iw, 
Sotiriou:2007zu} for clarifying discussions about the role of 
conformal frames in scalar-tensor gravity and the difference 
between different theories and different representations  of the 
same theory.}

Our aim is to generalize Hawking's results for black holes in 
Brans--Dicke theory to the more general class of actions of 
scalar-tensor gravity in eq.~(\ref{staction}), {\em i.e.}, to show  
that isolated black holes that are the end-state of 
collapse in  these theories are no different than in general 
relativity. So far, this has been shown only under the 
additional assumption of spherical symmetry, see 
Refs.~\cite{Mayo:1996mv,beken} and references therein.  Instead, we will not 
impose any symmetry assumption apart from stationarity, which 
simply reflects the condition that the black hole be 
the endpoint  of collapse. That is, we will only require 
spacetime to have a Killing vector which is timelike at 
infinity.

The field equation one derives from the 
action~(\ref{staction}) by varying with respect to 
$\hat{g}_{\mu\nu}$ and $\varphi$ are
\bea
\hat{R}_{\mu\nu}-\frac{1}{2} \hat{R} \hat{g}_{\mu\nu}\!\!& = &\!\!
\frac{\omega(\varphi)}{\varphi^2}\left(\hat{\nabla}_\mu\varphi\hat{\nabla}_\nu\varphi 
-\frac{1}{2}\hat{g}_{\mu\nu} \, \hat{\nabla}^\lambda\varphi\hat{\nabla}_\lambda\varphi\right) \nonumber\\
&&\nonumber\\
\!\!\!& + &\!\!\!\frac{1}{\varphi}\left(\hat{\nabla}_\mu\hat{\nabla}_\nu\varphi 
-\hat{g}_{\mu\nu}\hat{\Box}\varphi\right)-\frac{V(\varphi)}{2\varphi} 
\hat{g}_{\mu\nu}\,,\\
(2\omega+3)\hat{\Box}\varphi \!\!& = &\!\!-\omega'
\, \hat{\nabla}^\lambda\varphi\hat{\nabla}_\lambda\varphi+\varphi \, V'-2V  \,,
\eea
respectively, where  
$\hat{\Box}=\hat{\nabla}^\lambda\hat{\nabla}_\lambda$, a 
prime denotes differentiation with respect to the argument, 
and we have neglected the matter since we are interested in 
vacuum solutions. The condition that the black hole be isolated 
can be translated  into the requirement that it should be 
asymptotically flat, {\em i.e.},~at spatial infinity the 
metric should 
approximate Minkowski space and the scalar should go over 
to a constant $\varphi_0$. However, this can only be true if 
$V(\varphi_0)=0$ and
\be
\label{Vcond}
\varphi_0 \, V'(\varphi_0)-2V(\varphi_0)=0\,.
\ee

It might seem that our requirement for asymptotic flatness 
strongly restricts the type of theories which we can consider:  
it is quite common for instance (especially in cosmologically 
motivated models) for scalar-tensor theories to have 
$V(\phi_0)\neq0$, which would lead to an effective cosmological 
constant. However, one has to bear in mind that asymptotic 
flatness is clearly an idealization reflecting the 
requirement to consider an isolated black hole. Indeed, any  
realistic black hole is embedded in a cosmological 
background. To the extent that it is reasonable to expect 
this cosmological background not to affect local physics, which 
is the very assumption under which one effectively neglects the 
stress-energy tensor of the cosmic fluid in the first place in 
such considerations, we should be able to safely neglect a 
non-vanishing $V(\varphi_0)$ as well. Formally, this can be 
achieved by shifting the potential by a suitable constant which 
one considers part of the stress-energy tensor of the cosmic 
fluid. 

Based on this consideration, for what concerns 
us here, we restrict our attention to theories for which 
$V(\varphi_0)=0$ for concreteness, but we regard this condition 
 as physically non-essential.  On the other hand, the 
condition in  eq.~(\ref{Vcond}) appears to be more fundamental. 
It essentially reflects the fact that, if the 
potential for  the scalar has no extrema or is not bound, 
there cannot be constant $\varphi$ solutions. 

Provided that the aforementioned conditions hold, it is evident  
that vacuum solutions of scalar-tensor gravity with 
$\varphi=\varphi_0$ will be solutions of general relativity. In 
the remainder of this paper 
we will argue that the only stationary, asymptotically flat 
black hole solutions {of scalar-tensor gravity are indeed 
solutions with $\varphi=\varphi_0$.
 
Instead of using the Jordan frame, one can use a different 
conformal frame, which can prove very advantageous for 
certain calculations. The most commonly used  alternative is the {\em Einstein frame}, which is 
defined as the frame in which the scalar couples minimally to 
gravity (and, therefore, non-minimally to matter). The 
conformal  transformation $g_{\mu\nu}=\varphi 
\, \hat{g}_{\mu\nu}$, together with the scalar field redefinition
\be
d\phi=\sqrt{\frac{2\omega(\varphi)+3}{16\pi}} \, \frac{d\varphi}{\varphi}\,,
\ee
brings the action~(\ref{staction}) to the form
\bea
\label{stactionein}
S_{\rm st}&=&\int d^4x \sqrt{-g} \Big(\frac{R}{16\pi}-\frac{1}{2} \nabla^\mu \phi \nabla_\mu \phi\nn\\&&\qquad\qquad\qquad-U(\phi)+L_m(\hat{g}_{\mu\nu},\psi)\Big)\,,
\eea
where  $U(\phi)=V(\varphi)/\varphi^2$ and $g_{\mu\nu}$ is 
(per definition) the {\em Einstein frame metric}. $R$ is the 
Ricci scalar of that metric, $g$ is its determinant, and 
$\nabla_\mu$ is the corresponding covariant derivative. 
Note that $\hat{g}_{\mu\nu}$ appears in $L_m$, effectively 
signaling the non-minimal coupling between matter and 
$\phi$.

Here we are interested in black hole solutions, which 
are vacuum solutions, and so we neglect  again $L_m$. 
It becomes immediately apparent that, in this 
case, we remain  with only one free function specifying the 
theory within the class, $U(\phi)$ (or $V(\varphi)$).
The field equations that one derives by varying the 
action~(\ref{stactionein}) with respect to $g_{\mu\nu}$ and 
$\phi$ are 
\bea
\label{field1}
R_{\mu\nu}-\frac{1}{2} R g_{\mu\nu}&=&8\pi\,T_{\mu\nu}^\phi\,,\\
\label{field2}
\Box \phi&=& U'(\phi)\,,
\eea
respectively, where
\be
T_{\mu\nu}^\phi=\nabla_\mu \phi \nabla_\nu \phi 
-\frac{1}{2}g_{\mu\nu}\nabla_\lambda \phi \nabla^\lambda \phi 
-U(\phi)g_{\mu\nu}\,.
\ee

Given that the conformal factor in the transformation that 
relates the Einstein and the Jordan frame is just $\varphi$, 
symmetries remain unaffected. That is, in the Einstein frame one 
will still have a Killing vector $\xi^\mu$ which is timelike at 
infinity. At the same time, asymptotically flat solutions are 
mapped into asymptotically flat solutions (provided that 
$\omega(\varphi)\neq-3/2$ and does not diverge\footnote{When 
$\omega=-3/2$ the scalar does not have dynamics and the theory 
reduces to general relativity in vacuo. This theory is 
equivalent to Palatini $f(R)$ gravity   
\cite{Flanagan:2003rb,Sotiriou:2006hs}.}). Finally, it is easy 
to verify that the condition~(\ref{Vcond}) corresponds to 
$U'(\phi_0)=0$ and that $V(\varphi_0)=0$ implies $U(\phi_0)=0$. 
These considerations provide the full set-up of the problem 
in the Einstein frame. 

The main reason for one to use the Einstein frame is that in 
this frame $T_{\mu\nu}^\phi$ does not contain second derivatives of the scalar. It can, therefore, satisfy the Weak Energy Condition \cite{Wald}, with the mild assumption that $U$ (or $V$ in the Jordan frame) does not become overly negative.
It has been shown in 
Ref.~\cite{hawking1} that stationary, asymptotically flat 
spacetimes which satisfy this energy condition are necessarily 
axisymmetric and, therefore, there is a second Killing vector 
$\zeta^\mu$ which is spacelike at spatial infinity. 

Now let ${\cal S}_1$ be a partial Cauchy hypersurface for 
$\bar{J}^+( \mathscr{I}^-)\cap\bar{J}^-( \mathscr{I}^+)$ (the 
intersection of the topological closure of the causal future of 
past null infinity with the topological closure of the causal 
past of future null infinity) and ${\cal S}_2$ a second Cauchy 
hypersurface obtained by moving each point of ${\cal S}_1$ a 
unit parameter distance along the integral curves of $\xi^\mu$. 
Consider the volume ${\cal V}$ bounded by a portion of the black 
hole horizon, the two Cauchy hypersurfaces and a timelike 
3-surface at infinity. Assume that the scalar is not sitting at 
the minimum of its potential everywhere.\footnote{$U'$ can be 
zero at a point, along a curve, or even on a 
hypersurface because the contributions to the integral from 
these lower-dimensional sets vanish.} Multiplying both 
sides of eq.~(\ref{field2}) by $U'$ and integrating over 
the 
volume ${\cal V}$ one obtains
\begin{equation}
\int_{\cal V} d^4 x\sqrt{-g}\, U'(\phi)\Box \phi= \int_{\cal V} 
d^4 x\sqrt{-g}\, U'^2(\phi).
\end{equation}
One can rewrite this equation as
\begin{eqnarray}
\int_{\cal V} d^4 x\sqrt{-g}\, \big[U''(\phi)\nabla^\mu \phi\nabla_\mu \phi + U'^2(\phi)\big]=\nonumber\\=\int_{\partial{\cal V}} d^3 x\sqrt{|h|}\, U'(\phi)\nabla_\mu\phi \,\,n^\mu,
\end{eqnarray}
where $\partial{\cal V}$ denotes the boundary of ${\cal V}$, 
$n^{\mu}$ is the normal to the boundary, and $h$ is the 
determinant of the induced metric $h_{\mu\nu}$ on the 
boundary. Now split the boundary into its constituents. The black 
hole horizon part will not contribute to the integral because 
the normal to the horizon is a linear combination of the two 
Killing vectors  
$\xi^\mu$ and $\zeta^\mu$, which generate the 
symmetry that the scalar respects, so $\xi^\mu 
\nabla_\mu\phi = \zeta^\mu \nabla_\mu\phi=0$. The integrals over 
the two Cauchy hypersurfaces will have opposite values and cancel 
each other. Finally, asymptotic flatness requires that the 
integral vanishes over the timelike surface at infinity, as 
$\phi\to \phi_0$. Therefore, one obtains
\begin{eqnarray}
\label{feq}
\int_{\cal V} d^4 x\sqrt{-g}\, \big[U''(\phi)\nabla^\mu \phi\nabla_\mu \phi + U'^2(\phi)\big]=0.
\end{eqnarray}
$U'^2(\phi)$ is manifestly positive or zero. Since the gradient 
of the scalar has to be orthogonal to both Killing vectors, it 
can only be spacelike or zero, which implies that $\nabla^\mu 
\phi\nabla_\mu \phi$ is positive  or zero as well. However, as 
long as 
\begin{equation}
\label{U''cond}
U''(\phi)\geq 0,
\end{equation}
eq.~(\ref{feq}) cannot be satisfied unless $\phi=\phi_0$ and 
$U'(\phi_0)=0$. 

From this one can conclude that, provided that 
the condition~(\ref{U''cond}) holds, the only stationary, 
asymptotically flat black hole solutions are solutions of general 
relativity. The condition~(\ref{U''cond}) has a 
straightforward physical interpretation: it is a local linear 
stability condition for the scalar, {\em i.e.},~if $U''< 0$ at a 
neighborhood of a spacetime point, then the 
scalar is unstable there. Therefore, in theories with multiple 
extrema in the potential, there can also exist stationary, 
asymptotically flat black hole solutions which are not solutions 
of general relativity, but they will be linearly unstable.

Let us now discuss the subtleties and limitations of the proof 
just presented. First of all, clearly our approach hinges on the 
assumption that $V(\varphi)\neq 0$ (or equivalently $U(\phi)\neq 
0$) in the first place. However, when $V(\varphi)=0$, Hawking's 
original proof for Brans--Dicke theory applies unmodified to 
scalar-tensor theories with $\omega=\omega(\varphi)$, provided 
that $\omega (\varphi) \neq -3/2$ and does not diverge. This is 
simply because, in the absence of matter, $\omega$ can be 
absorbed in the scalar redefinition in the Einstein frame.

Theories where $\omega$ can diverge are not covered by our 
proof. Also, for all theories, there is the possibility to have 
solutions which are different from general relativity if 
$\varphi$ diverges somewhere in spacetime, as this would signal a 
breakdown of the conformal transformation from the Jordan  to the 
Einstein frame. For example, such a maverick solution where 
$\varphi$ diverges on the horizon has been found for a 
conformally coupled scalar \cite{maverick} (though it is linearly 
unstable \cite{unstable}). It is worth mentioning that these 
restrictions on $\omega$ and $\varphi$ do not apply if the scalar 
field is assumed to be minimally coupled in the Jordan frame  
(then Jordan and Einstein frame coincide), as for example in 
simple quintessence models. Lastly, as mentioned earlier, 
{metric $f(R)$ 
theories of gravity are equivalent to the special class of 
scalar-tensor theories with $\omega=0$ and a 
potential whose functional  form is related to the functional form of 
$f$ \cite{Teyssandier:1983zz,Barrow:1988xi,Wands:1993uu}.
As such, they are covered by our proof.

In our approach we have completely neglected the matter
sector of the theory. However, it is straightforward to verify 
that allowing the presence of conformally invariant matter  will not  
affect the outcome. The requirement  of 
conformal invariance guarantees that, given that matter does not 
couple to $\varphi$ in the Jordan frame, it will also not  
couple to $\phi$ in the  Einstein frame. Therefore, our 
results are actually  applicable not only to vacuum, 
but also to electro-vacuum black holes.

Finally, let us re-examine our two assumptions, namely that 
the  black holes should be stationary and asymptotically flat. 
The first assumption, as we have repeatedly mentioned, reflects the  requirement that the black holes represent 
quiescent objects which are the final state of gravitational 
collapse. The second assumption stems from the requirement that 
the black holes be isolated objects. Imposing asymptotic flatness 
forces one to neglect the contribution of a(n) (effective) 
cosmological constant,  as well as of the background cosmology in 
which the black hole will  realistically be embedded. However, to 
the extent that one can trust that local physics is not 
affected by such contributions, and for  black holes whose 
characteristic scale is much smaller that the 
Hubble scale, we  expect asymptotic flatness to be an 
excellent approximation  (in fact it is the standard assumption 
in black hole physics and for solutions describing stars).

To summarize, we  have considered the fairly general class of 
scalar-tensor theories of gravity (which also includes $f(R)$ 
gravity) and we have shown that isolated black holes which are 
the end-state of collapse are solutions of this class if and 
only if they  are solutions of general relativity, unless: 
(i)~they are linearly unstable, or (ii)~the scalar is 
non-minimally coupled to gravity and diverges somewhere in 
spacetime, or (iii)~the scalar violates the Weak Energy Condition in the Einstein frame.

{\em Acknowledgments:}
We would like to thank Stefano Liberati, Matt Visser and Vincenzo 
Vitagliano for stimulating discussions. T.P.S. acknowledges partial financial support provided under the "Young SISSA Scientists' Research Projects" scheme 2011-2012, promoted by the International School for Advanced Studies (SISSA), Trieste, Italy.
VF thanks Bishop's University and the Natural Sciences and 
Engineering Research Council of Canada (NSERC) for financial
support.



\end{document}